\documentclass[10pt,twocolumn,letterpaper]{article}

\usepackage{cvpr}
\usepackage{times}
\usepackage{epsfig}

\usepackage{colortbl}
\usepackage{times}
\usepackage{graphicx}
\usepackage{amsmath,amssymb,amsfonts}
\usepackage{multirow}
\usepackage{bigstrut}
\usepackage{array}
\usepackage{url}
\usepackage{multirow}

\usepackage{float}

\usepackage[breaklinks=true,bookmarks=false]{hyperref}

\cvprfinalcopy 

\begin{document}

\author{S. M. Kamrul~Hasan$^{1}$\thanks{Research reported in this publication was supported by the National Institute of General Medical Sciences of the National Institutes of Health under Award No. R35GM128877 and by the Office of Advanced Cyber infrastructure of the National Science Foundation under Award No. 1808530.},
~Cristian A.~Linte$^{1,2}$\\
$^{1}${Center for Imaging Science, $^{2}$Biomedical Engineering}\\
{Rochester Institute of Technology, Rochester, NY}\\
{\tt\small \{sh3190, calbme\}@rit.edu}}

\title{L-CO-Net: Learned Condensation-Optimization Network for Clinical Parameter Estimation from Cardiac Cine MRI }

\maketitle
\begin{abstract}
In this work, we implement a fully convolutional segmenter featuring both a learned group structure and a regularized weight-pruner to reduce the high computational cost in volumetric image segmentation. We validated our framework on the ACDC dataset featuring one healthy and four pathology groups imaged throughout the cardiac cycle. Our technique achieved Dice scores of 96.80\% (LV blood-pool), 93.33\% (RV blood-pool) and 90.0\% (LV Myocardium) with five-fold cross-validation and yielded similar clinical parameters as those estimated from the ground-truth segmentation data. Based on these results, this technique has the potential to become an efficient and competitive cardiac image segmentation tool that may be used for cardiac computer-aided diagnosis, planning and guidance applications.
\end{abstract}

\section{Introduction}
\label{sec:intro}

The emerging success of Convolutional Neural Networks (CNNs) in solving high-level computer vision tasks can be utilized to develop machine learning tools that are capable of learning hierarchical features in an end-to-end manner \cite{hasan2019u, kirillov2019panoptic}. Motivated by the superior performance of deep learning, the medical imaging research community has also embraced
the implementation of deep learning-based approaches for medical image segmentation \cite{ronneberger2015u}, as a precursor task for clinical parameter estimation \cite{suinesiaputra2018fully}. However, image segmentation in clinical settings still requires more accuracy and precision, with even minimal segmentation errors being unacceptable. 
In the context of cardiac image segmentation, fully convolutional networks (FCNs) have become well established, thanks to their per pixel prediction capabilities. An example of such an application is the segmentation of various cardiac structures from MR images \cite{tran2016fully}. Similarly, Bai {\it et al.} \cite{bai2018automated} reported improved accuracy and robustness of the ventricles and atria segmentation by using a modified FCN architecture.


\begin{figure}[t!]
\includegraphics[width=1.0\linewidth]{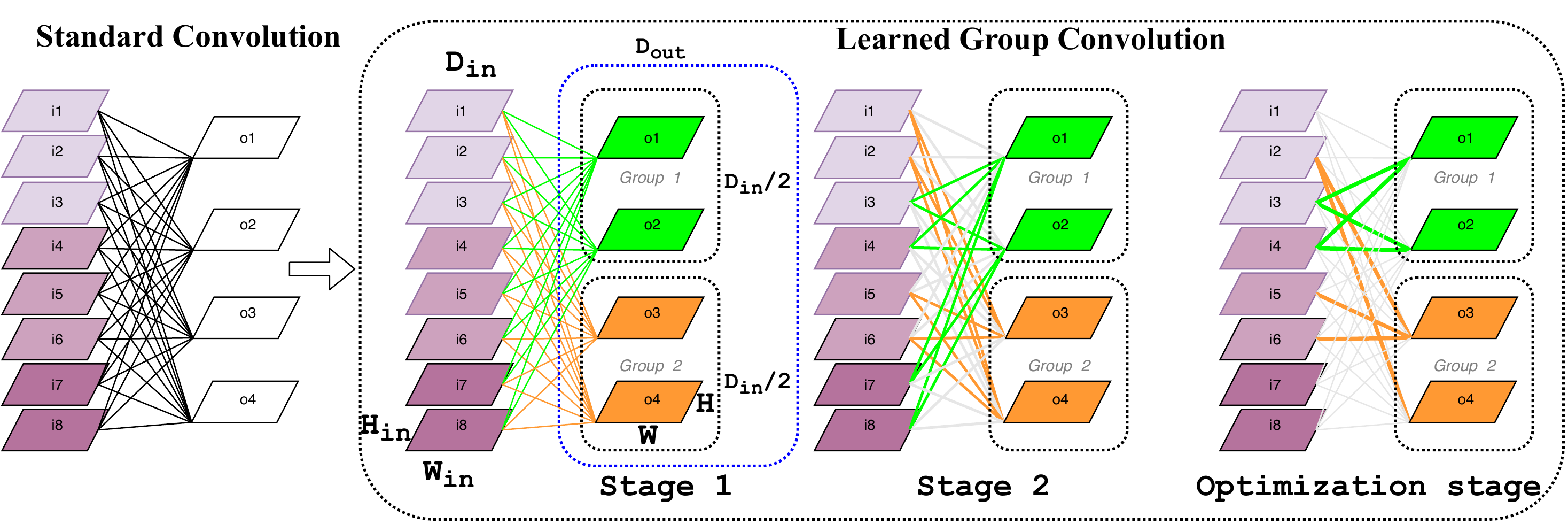}
  \caption{Illustration of Learned Group Convolution (LG-Conv) block through multi-stage system in \textit{L-CO-Net}.}
\label{fig:LGCONV}
\end{figure}


The formulation and integration of various regularization techniques has been a growing strategic trend to improve the generalization performance of neural networks. One such particularly compelling approach is the use of Dropout at the training stage of a neural network. However, the accuracy of a trained deep network will not be severely improved by dropping out a majority of connections at the training stage and hence current research efforts have been focused on the use of deep model compression tasks, including weight pruning \cite{ye2018progressive}, weight decay \cite{zhang2018three}, and knowledge distillation \cite{hinton2015distilling}. In this work, we also utilize a weight-pruning‐based network regularization approach.

Weight-pruning has aroused much research attention due to its faster inference with minimal loss in accuracy. Gao Huang {\it et al.} demonstrated the use of weight-pruning technique in a group-convolution setting, where a DenseNet type architecture can learn more sparse information during the training process and prune redundant connections between convolution layers \cite{huang2018condensenet}.



\begin{figure*}[t!]
\includegraphics[width=1.0\linewidth]{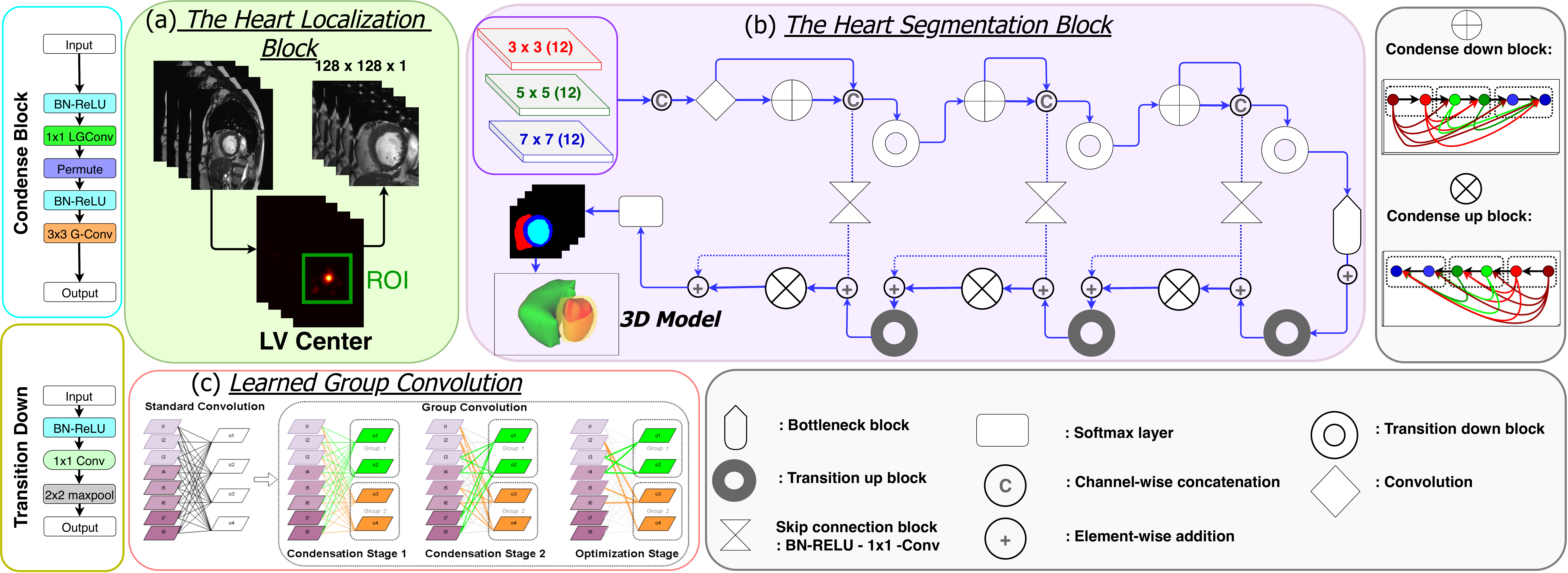}
  \caption{Illustration of \textit{L-CO-Net} framework: \textbf{(a)} ROI detection around LV-RV; \textbf{(b)} Segmentation block consisting of a decoder and an encoder where each condense block (CB) consists of 3 Layers with a growth rate of k = 16. The transformations within each CB and the transition-down block are labeled with a cyan and yellow box, respectively.}
\label{fig:L-CO-Net}
\end{figure*}


In this work, we propose to use the concept of learned-group convolution and weight-pruning technique in a fully convolutional setting to segment the left and right ventricle blood-pool and left ventricle Myocardium from end-diastolic and end-systolic cardiac MR images in a more accurate and more efficient manner. To assess the performance of this proposed framework, we compare our results (Dice score, Hausdorff distance, and clinical parameters) to those obtained using five other segmentation architectures on the Automatic Cardiac Diagnosis Challenge (ACDC) dataset. Lastly, we show that the proposed learned-group convolution and weight-pruning technique improve the segmentation performance \cite{hasan2020condenseunet}, as well as the estimation of clinical cardiac indices in cine MR slices.



\section{Methodology}
\label{sec:method}
To tackle the task of precise and rapid heart chamber detection and segmentation in cine MR images, we propose a specifically designed network architecture ---  \textit{learned condensation-optimization network (L-CO-Net)}, shown in Figure \ref{fig:L-CO-Net}. Our proposed \textit{L-CO-Net} framework substitutes the concept of both standard convolution and group convolution (G-Conv) with learned group-convolution (LG-Conv). While the standard convolution needs an increased level of computation, i.e. $\mathcal{O}(I_i$ x $O_o)$, and concurrently, the pre-defined use of filters in each group convolution \cite{xie2017aggregated} restricts its representation capability, these aforementioned problems are mitigated by introducing \textit{LG-Conv} that learns group convolution dynamically during training through a multi-stage scheme. Before training, the input channels and filters are split into equally sized $M$ groups denoted as $I^k$ = \{$I_1^k$, $I_2^k$, ... , $I_M^k$ \} and  $F^k$ = \{$F_1^k$, $F_2^k$, ... , $F_M^k$\}, where $I_i^k$ is the $i^{th}$ feature map of $k^{th}$ layer. The output of this group convolution layer is formulated as $I^{k+1}$ = $[ F_1^k \otimes I_1^k, F_2^k \otimes I_2^k, ... , F_M^k \otimes I_M^k ]$ = 
$[\{ f_{11}^k \ast i_{11}^k, f_{12}^k \ast i_{12}^k, ... , f_{1N}^k \ast i_{1h}^k $\},
$\{ f_{21}^k \ast i_{21}^k, f_{22}^k \ast i_{22}^k, ... , f_{2N}^k \ast i_{2h}^k $\},
.... , $\{ f_{M1}^k \ast i_{M1}^k, f_{M2}^k \ast i_{M2}^k, ... , f_{MN}^k \ast i_{Mh}^k \}]$, where $I^k$ = \{$i_1^k$, $i_2^k$, ... , $i_h^k$ \}, $F^k$ = \{$f_1^k$, $f_2^k$, ... , $f_N^k$\}, $h$ is the number of channels, and $N$ is the number of filters. Since each group has its own weights, they can select their own set of relevant input features, assisting the system to predict most relevant features at the relevant connections. This multi-stage pipeline consists of \textit{multi-condensation} stages followed by the \textit{optimization} stage. In the first half of the pipeline, training is initiated by calculating the magnitude of the weights for each incoming feature, which are then averaged. After that, the low-magnitude weighted column is screened out from the features. Thus, a fraction of $(C -1)/C$ is truncated after each of the $C-1$ condensing stages.

The second part of the pipeline is where all training occurs. This stage is focused on finding the optimal permutation connection that will share a similar sparsity pattern, to mitigate any negative effects on accuracy induced by the pruning process (Figure \ref{fig:LGCONV}). As mentioned by Huang {\it et al.}, in their paper \cite{huang2018condensenet}, both the $L_1$ and $L_2$ regularization methods are efficient for solving the overfitting problem, but they do not perform well for network optimization. To address this limitation, we introduce an efficient regularizer referred to as group lasso (GL), which is a natural generalization of the standard lasso (least absolute shrinkage and selection operator) objective \cite{friedman2010note}. Additionally, the GL regularizer encourages group-level sparsity at the factor level by forcing all outgoing connections from a single neuron (corresponding to a group) to be either simultaneously zero or not.


\begin{figure*}[t!]
\includegraphics[width=1.0\linewidth]{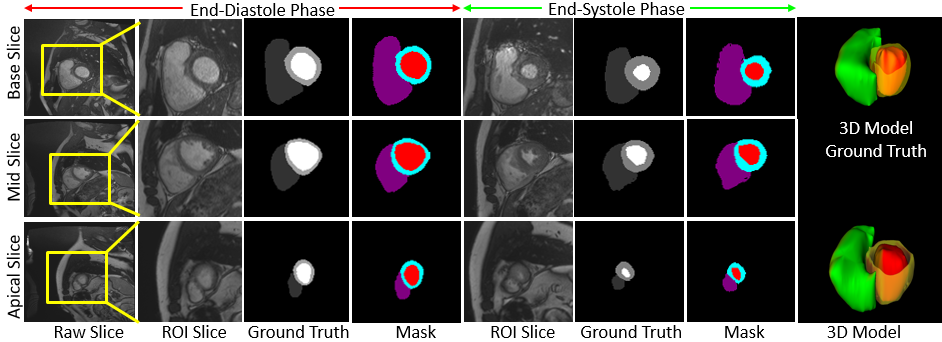}
  \caption{Representative ED and ES frames segmentation results of a complete cardiac cycle from the base (high slice index) to apex (low slice index) showing RV blood-pool, LV blood-pool, and LV-Myocardium in purple, red, and cyan respectively.}
\label{fig:results}
\end{figure*}


\subsection{Heart Localization}
To reduce computational complexity and improve accuracy, a Fourier transform-based method proposed by Xiang Lin {\it et al.} \cite{lin2006automated} is used to automatically detect and extract a region of interest (ROI) that encompasses the LV and RV. The motivation for using the Fourier transform is that LV and RV are the only large moving structures in the thorax and move at the same frequency, dictated by the heart rate. Therefore, the pixel intensity changes over time between the LV blood-pool and the LV-myocardium, whereas the change in pixel intensity is almost static at the boundary. We enhanced the LV and RV regions by computing the Fourier transform for each slice and retaining only the first harmonic. Moreover, since the shape of the LV is circular in nature, we also used the circle Hough transform introduced by Ilkay Oksuz {\it et al.} \cite{oksuz2019automatic} to identify the center and radius of the ROI of the LV and RV. We then generated a bounding-box and used it to crop the ROI from the image.

\subsection{Heart Segmentation}

This block consists of both an encoder and a decoder path, where the encoder path has an input image size of $128 \times 128$, and three condense blocks (CBs) with feature map size of $\{128^2, 54^2, 32^2\}$. We employ separable convolution with different filter sizes in the initial layers and then stacked them together, as inspired by the Xception network. 

We introduced a novel skip connection block which is computationally and memory-efficient (Figure \ref{fig:L-CO-Net}). The decoder is symmetrical to the encoder consisting of three blocks, comprised of $3 \times 3$ transposed convolutions CBs, and a soft-max layer in the last layer for generating the image mask. The concatenation in skip-layer has been replaced by an element-wise addition operation to mitigate the problem of the feature-map explosion. We employ a number of layers per block as 2, 3, 4, 5, 4, 3, 2 with 32 initial feature maps, 3 max-pooling layers, a growth rate of k = 16, group/condense block = 4, and condensation factor, C = 4 (Figure \ref{fig:L-CO-Net}). The weights are updated during back-propagation operation by minimizing the dual loss function, $\mathcal{L}_{Total}$:

\begin{equation} \label{eq:1}
\mathcal{L}_{Total} = \alpha.\mathcal{L}_{Entropy}(A, E) + \beta. (1-\mathcal{L}_{Dice}(A, E))
\end{equation}

\noindent
where $\mathcal{L}_{Entropy}$ is the weighted cross-entropy loss and $\mathcal{L}_{Dice}$ is the dice loss. The parameter $\alpha$ varies between $0$ and $1$ and $\beta$ $=$ $1-\alpha$. $A$ be the training samples and $E$ be the weights. The first term, $\mathcal{L}_{Entropy}$ in equation \ref{eq:1} is used to calculate the weight map from the reference classes and labels, where $L$ and $|V|$ are the set of all reference classes and voxels in the training set, respectively in equation \ref{eq:2}.
\begin{equation} \label{eq:2}
\begin{split}
& \mathcal{L}_{Total}=\alpha.~ [-\sum_{a_{i}\in A}\{\sum_{l\in L}\frac{scale*|V|}{class_{freq}}~+ \\
& ~~~~~~~~~~~~\sum_{l\in L}\frac{edge_{scale}*|V|}{edge_{freq}}\}~{log}~(p({r_i}|{a_i} ;E))]~+\\
& ~~~~~~\beta. ~[1- \frac{\sum_{l\in L}\frac{|B|}{|B_l|}(\sum_{a_i\in A}~p({r_i}|{a_i} ;E)G({a_i})+\epsilon)}{\sum_{l\in L}\frac{|B|}{|B_l|}(\sum_{a_i\in A}~p({r_i}|{a_i} ;E)+G(a_i)+\epsilon )}] \\
\end{split}
\end{equation}

\noindent
Let $r_i$ be the label of the reference class corresponding to voxel $a_i \in A$. $|B|$ represents the number of pixels in a mini-batch and $|B_l|$ represents the number of pixels in each class $l \in L$. The term $\epsilon$ is used to prevent division by 0, when one of the sets is empty. The total loss, $\mathcal{L}_{Total}$ is minimized via the Adam optimizer and evaluated by dice scores associated with clinical indices i.e. ejection fraction and myocardial mass etc.


\subsection{Imaging Data}
For this study, we used the Automated Cardiac Diagnosis Challenge (ACDC) dataset\footnote{https://www.creatis.insa-lyon.fr/Challenge/acdc/databases.html}, consisting of short-axis cardiac cine-MR images acquired for 100 patients divided into 5 subgroups: normal (NOR), myocardial infarction (MINF), dilated cardiomyopathy (DCM), hypertrophic cardiomyopathy (HCM), and abnormal right ventricle (ARV), available through the 2017 MICCAI-ACDC challenge \cite{bernard2018deep} which are splitted into 70\% training and 15\% validation set.

\begin{table}[h!]
\scriptsize
\caption {Quantitative evaluation of the segmentation results in terms of Mean Dice score (\%) with Hausdorff distance(in mm), no. of parameters ($\times 10^6$), and the clinical indices evaluated on the ACDC dataset for LV, RV blood-pool and LV-myocardium compared across several best performing networks, including \textit{L-CO-Net}. The statistical significance of the results for L-CO-Net model compared against five other baseline models are represented by $*(p < 0.05)$ and $**(p < 0.01)$.}
\begin{center}
\begin{tabular}{p{1.14cm}>{\centering\arraybackslash}m{0.7cm}>{\centering\arraybackslash}m{0.7cm}>{\centering\arraybackslash}m{0.7cm}>{\centering\arraybackslash}m{0.7cm}>{\centering\arraybackslash}m{0.7cm}>{\centering\arraybackslash}m{0.8cm}}

\hline
\multirow{2}{1.6cm} & \multicolumn{6}{c}{End Diastole (ED)}\\ 
\cline{2-7}
& UNet  & DCN & MUNet& MNet  & DNet & \textbf{L-CO-Net}  \\
\hline
Dice [LV]& 95.0(8.2)  & 96.0(7.5)  & 96.3(6.5)  & 96.1(7.7) & 96.4(8.1)  & \underline{*96.8}(7.9) \\
Dice [Myo] & 88.2(9.8)  & 87.5(11.1) & 89.2(8.7)  & 87.5(9.9) & {88.9}(9.8) & \underline{*89.5}(8.9)  \\
Dice [RV] & 91.1(13.5) & 92.8(11.9) & 93.2(12.7) & 92.9(12.9) & \underline{93.5}(14.0) &  93.3(11.2)  \\
\hline
\end{tabular}
\label{Tab:Dice}
\end{center}
\end{table}


\begin{table}[h!]
\scriptsize
\vspace{-6.5mm}
\begin{center}
\begin{tabular}{p{1.15cm}>{\centering\arraybackslash}m{0.7cm}>{\centering\arraybackslash}m{0.7cm}>{\centering\arraybackslash}m{0.7cm}>{\centering\arraybackslash}m{0.7cm}>{\centering\arraybackslash}m{0.7cm}>{\centering\arraybackslash}m{0.7cm}}
\hline
\multirow{2}{1.4cm} & \multicolumn{6}{c}{End Systole (ES)}\\ 
\cline{2-7}
\hline
Dice [LV]& 90.0(10.9) & 91.0(9.6) & 91.1(9.2) & 91.5(7.1) & 91.7(9.0) & \underline{**95.1}(6.4)\\
Dice [Myo] & 89.7(11.3) & 89.4(10.7) & 90.1(10.6)  & 89.5(8.9)  & 89.8(12.6)  & \underline{*90.0}(8.9) \\
Dice [RV] & 81.9(18.7) & 87.2(13.4) & 88.3(14.7) & 88.5(11.8) &\underline{87.9}(13.9)& 87.4(11.9) \\
\hline
\end{tabular}
\end{center}
\vspace{-3.5mm}
\end{table}


\section{Results}
The proposed architecture was evaluated on the MICCAI STACOM 2017 ACDC dataset in a stratified five-fold cross validation. Figure \ref{fig:results} shows segmentation results and the ground truth masks for both 2D and 3D cases. Table \ref{Tab:Dice} summarizes the comparison results, which show that our proposed model significantly improved the segmentation performance against several state-of-the-art multi-class segmentation techniques \cite{bernard2018deep} in terms of Dice metrics, Hausdorff distance, and clinical parameters. Our proposed L-CO-Net architecture achieved Dice score (Hausdorff distance) of $96.8\%(7.9mm)$ and $95.1\%(6.4mm)$ for the LV blood-pool, $89.5\%(8.9mm)$ and $90.0\%(8.9mm)$ for the LV-Myocardium and $93.3\%(11.2mm)$ and $87.43\%(11.9mm)$ for the RV blood-pool in end-diastole and end-systole, respectively. 

The predicted segmentation was subsequently used to compute the clinical parameters using Simpsons method and the agreement between the ground truth and the automatic is reported using correlation statistical analysis by mapping the predicted volumes of the testing set onto the ground truth volumes of the training set. As illustrated in Table \ref{Tab:correlation} the agreement between our method's prediction and ground truth is high, characterized by a Pearson's correlation coefficient (rho) of $0.997 (p < 0.01)$ for LV-EF, $0.998$ for LV-EDV and $0.993 (p < 0.1)$ for Myo-mass. There was a slight over-estimation in the RV blood-pool segmentation also reflected in the clinical parameters estimation.

\begin{figure}[t!]
\begin{center}
\includegraphics[width=1.1\linewidth]{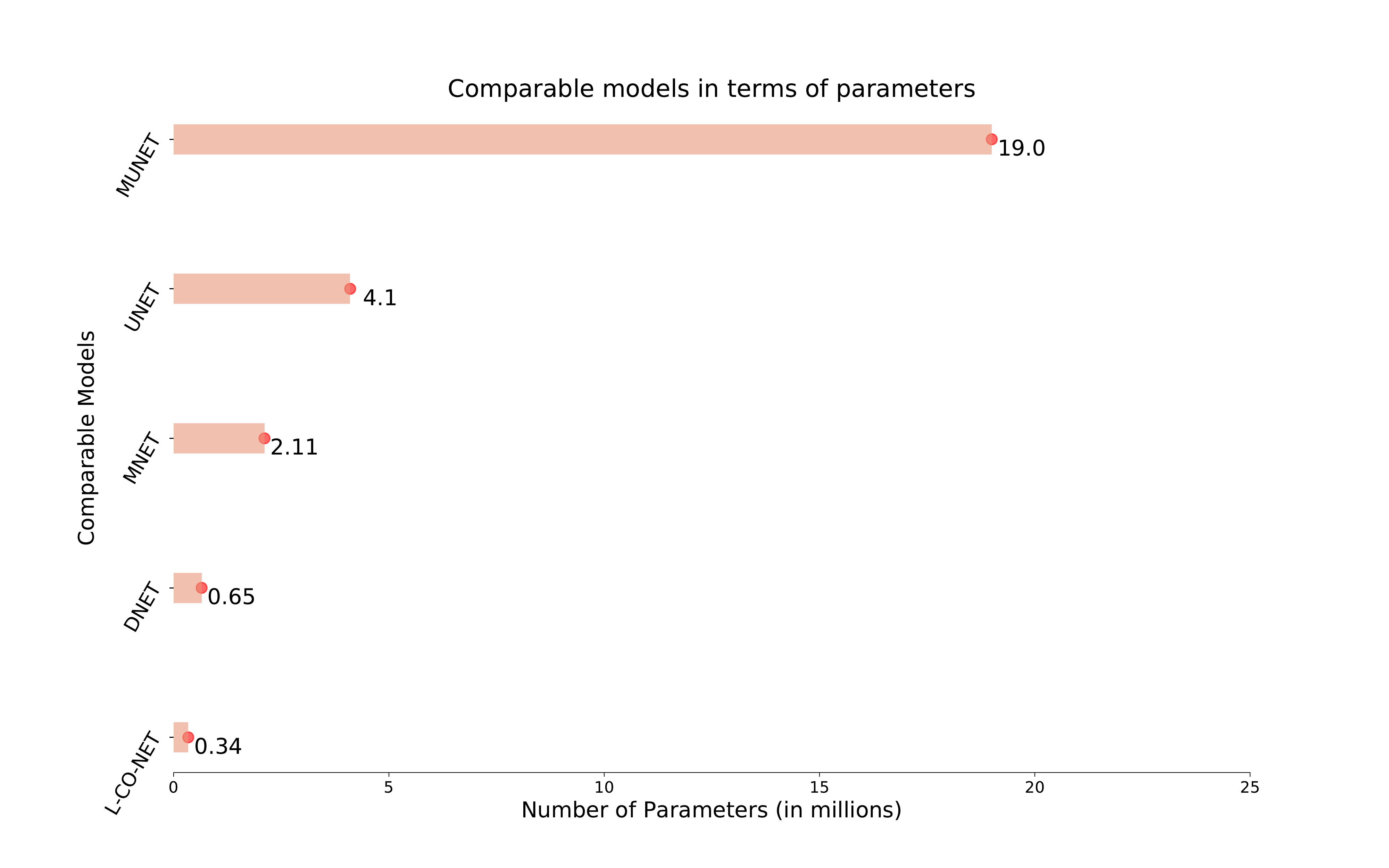}
\end{center}
\caption{Graphical comparison between clinical parameters estimated using L-CO-Net segmentation and same parameters estimated using the ground truth segmentation.}
\label{fig:corr}
\end{figure}
Figure \ref{fig:corr} shows a graphical comparison between the clinical parameters estimated from the cardiac features segmented via \textit{L-CO-Net} and the same homologous parameters estimated from the ground truth, manual segmentations, for both healthy volunteers and patients featuring various cardiac conditions. As shown, the clinical parameters estimated using our automatically segmented features show no significant difference from those estimated based on the ground truth, manually segmented features.



\begin{figure*}[ht]
\begin{center}
\includegraphics[width=0.5\linewidth]{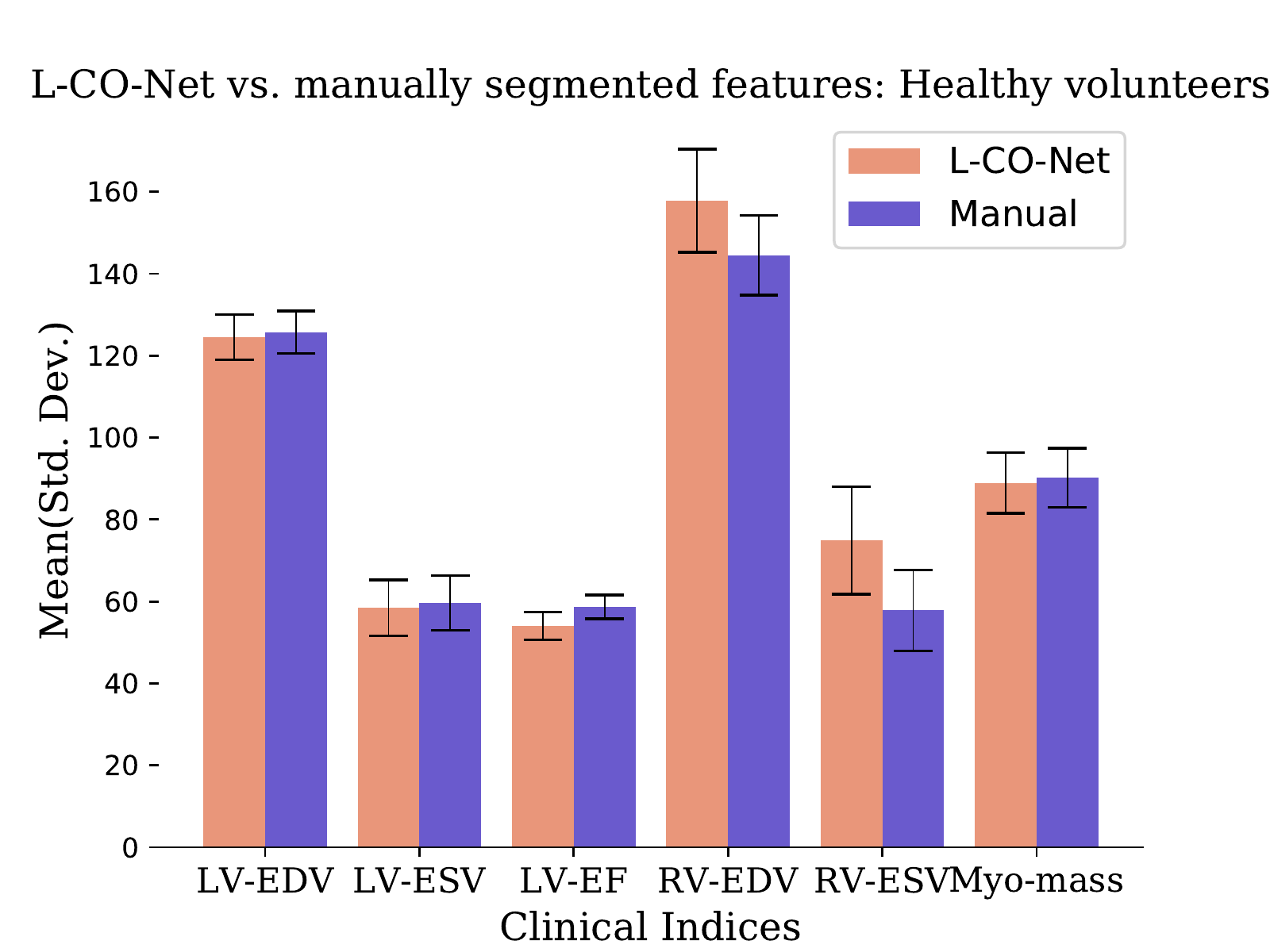}
\includegraphics[width=0.5\linewidth]{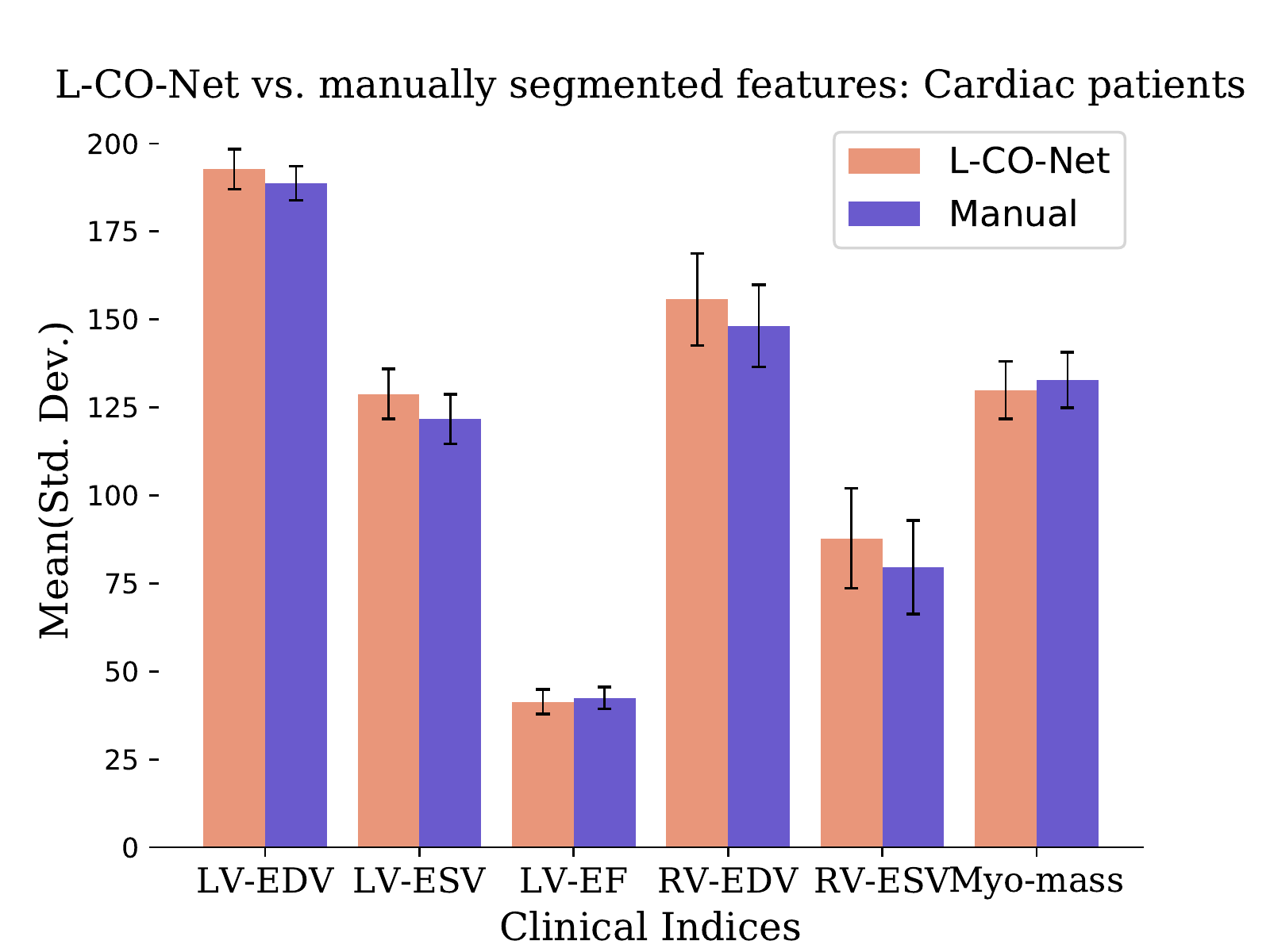}
\end{center}
\caption{Graphical comparison between clinical parameters estimated using L-CO-Net segmentation and same parameters estimated using the ground truth segmentation. EDV (in mL) = end-diastolic volume, ESV (in mL) = end-systolic volume, SV (in mL) = stroke volume, EF (\%) = ejection fraction MM (in gm) = myocardial mass}
\label{fig:corr}
\end{figure*}



\begin{table}[h!]
\scriptsize
\begin{center}
\caption{correlation between clinical parameters estimated using L-CO-Net segmentation and homologous parameters estimated from six other baseline segmentation methods ($\star (p < 0.1)$, $\star \star(p < 0.01)$).}
\begin{tabular}{p{1.09cm}>{\centering\arraybackslash}m{0.5cm}>{\centering\arraybackslash}m{0.5cm}>{\centering\arraybackslash}m{0.6cm}>{\centering\arraybackslash}m{0.6cm}>{\centering\arraybackslash}m{0.5cm}>{\centering\arraybackslash}m{0.5cm}>{\centering\arraybackslash}m{0.7cm}}

\hline
\multirow{2}{1.7cm} & \multicolumn{6}{c}{Correlation Coefficient}\\ 
\cline{2-8}
& UNet  & DCN & MUNet & MNet  & DNet & EUNet &\textbf{L-CO-Net}\\
\hline
LV EF & 0.987  & 0.988  & 0.988  & 0.989 & 0.989 & 0.991 & \underline{0.997}$\star \star$  \\
LV EDV & 0.997  & 0.993 &  0.995  & 0.993  & 0.997 & 0.997 & \underline{0.998}   \\
RV EF & 0.791 & 0.852 & 0.851 & 0.793 & 0.858 & \underline{0.901} &  0.869 \\
RV EDV & 0.945 & 0.980 & 0.977 & 0.986 & 0.982 & 0.988 &  \underline{0.988}   \\
Myo mass & 0.989 & 0.963 & 0.982 & 0.968 & 0.990  & 0.989 & {\underline{0.993}}$\star$  \\
\hline
\hline
\end{tabular}
{\raggedright DCN: Dilated Convolution Network, MUNet: Modified 3D UNet, MNet: Modified M-Net, DNet: DenseNet, EUNet\cite{isensee2017automatic}: Ensemble UNet, L-CO-Net: learned condensation-optimization Net.\par}
\label{Tab:correlation}
\end{center}
\end{table}

\noindent
In terms of performance, as summarized in Table \ref{Tab:Dice}, our proposed L-CO-Net segmentation framework entails roughly $340,000$ parameters, which represents more than 10 fold reduction from the UNet ($\sim4.1$ million parameters), 60 fold reduction from MUNet ($\sim19$ million parameters), and roughly 2 fold reduction from the most parameter-efficient method reported here - DNet ($\sim650,000$ parameters).

\section{Discussion and Conclusion}

In this paper, we propose a new memory-efficient architecture for accurate LV, RV blood-pool and myocardium segmentation, and clinical parameter quantification from breath-hold cine cardiac MRI. The capability of our network to learn the group structure allows multiple groups to re-use the same features via condensed connectivity. Moreover, the efficient weight-pruning methods leads to high computational savings without compromising segmentation accuracy. To the best of our knowledge, this is the first paper that presents a learned condensation-optimization approach for estimating clinical parameters from cardiac image segmentation in a fully convolutional setting. Our analysis across both healthy and abnormal patients indicated that the segmentation and estimated clinical parameters show no statistically significant difference from the ground truth manual segmentation and the inherently estimated clinical parameters.

Our proposed model outperforms several best methods according to dice scores, Hausdorff distances(HD), and clinical parameters,  achieving $96.8\%$ dice with $7.9mm$ HD for LV blood pool in ED and $95.1\%(6.4mm)$ in ES phase, which showed at least $0.41\%$ improvement in ED phase and $3.7\%$ improvement in ES phase over the current methods, as well as more than $6\%$ improvement over the traditional U-Net architecture. For LV-Myocardium segmentation, we achieved $89.5\%(8.9mm)$ in ED and $90.0\%(8.9mm)$ in ES, which showed at least $0.67\%$ improvement in ED and $0.22\%$ improvement in ES phase over the current methods, with at least a 10 fold reduction in the number of parameters. To improve the robustness of L-CO-Net framework, we used a low-level image pre-processing operation which serves as a precursor preliminary segmentation that narrows the capture range of the subsequent deep learning segmentation and parameter estimation. Our experiments show that L-CO-Net runs on the ACDC dataset using half $(50\%)$ the memory requirement of DenseNet and one-twelfth $(\sim 8\%)$ of the memory requirements of U-Net, while still maintaining excellent clinical accuracy. We observed that the segmentation results for RV have not improved significantly beyond those of the LV or myocardium.

An alternative solution for better segmentation of the RV would be to explore an additional slice refinement and slice misalignment correction as a future work.




{\small
\bibliographystyle{ieee_fullname}
\bibliography{root}
}

\end{document}